# ADVANCEMENTS IN BIG DATA PROCESSING IN THE ATLAS AND CMS EXPERIMENTS[1]

A.V. Vaniachine on behalf of the ATLAS and CMS Collaborations

*Argonne National Laboratory, 9700 S Cass Ave, Argonne, IL, 60439, USA*

**Abstract:** The ever-increasing volumes of scientific data present new challenges for distributed computing and Grid technologies. The emerging Big Data revolution drives exploration in scientific fields including nanotechnology, astrophysics, high-energy physics, biology and medicine. New initiatives are transforming data-driven scientific fields enabling massive data analysis in new ways.

In petascale data processing scientists deal with datasets, not individual files. As a result, a task (comprised of many jobs) became a unit of petascale data processing on the Grid. Splitting of a large data processing task into jobs enabled fine-granularity checkpointing analogous to the splitting of a large file into smaller TCP/IP packets during data transfers. Transferring large data in small packets achieves reliability through automatic re-sending of the dropped TCP/IP packets. Similarly, transient job failures on the Grid can be recovered by automatic re-tries to achieve reliable 6σ production quality in petascale data processing on the Grid.

The computing experience of the ATLAS and CMS experiments provides foundation for reliability engineering scaling up Grid technologies for data processing beyond the petascale.

## 1. Introduction

Today, various projects and initiatives are under way addressing the challenges of Big Data. For example, the data produced at the LHC and other advanced instruments present a challenge for analysis because of petascale data volumes, increasing complexity, distributed data locations and chaotic access patterns.

## 2. Big Data Processing at the LHC

To address petascale Big Data challenge, the LHC experiments are relying on the computational infrastructure deployed in the framework of the Worldwide LHC Computing Grid. Following Big Data processing on the Grid, more than 8000 scientists analyze LHC data in search of discoveries. Culminating the search, a seminar at CERN on July 4 presented the results of the Higgs boson search at the LHC. Figure 1 shows the significance of the search achieved by the ATLAS experiment in combination of several Higgs decay channels. A standard for discovery – the five sigma significance of the result – corresponds to the $3 \cdot 10^{-7}$ probability of the background fluctuation to mimic the Higgs signal (local $p_0$ value).

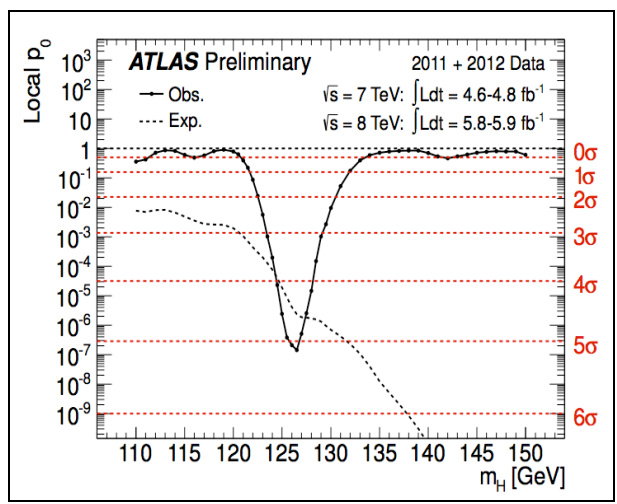

**Figure 1:** The Higgs boson discovery culminated many years of search for new physics phenomena driven by the Big Data revolution in Grid computing technologies [1].

---

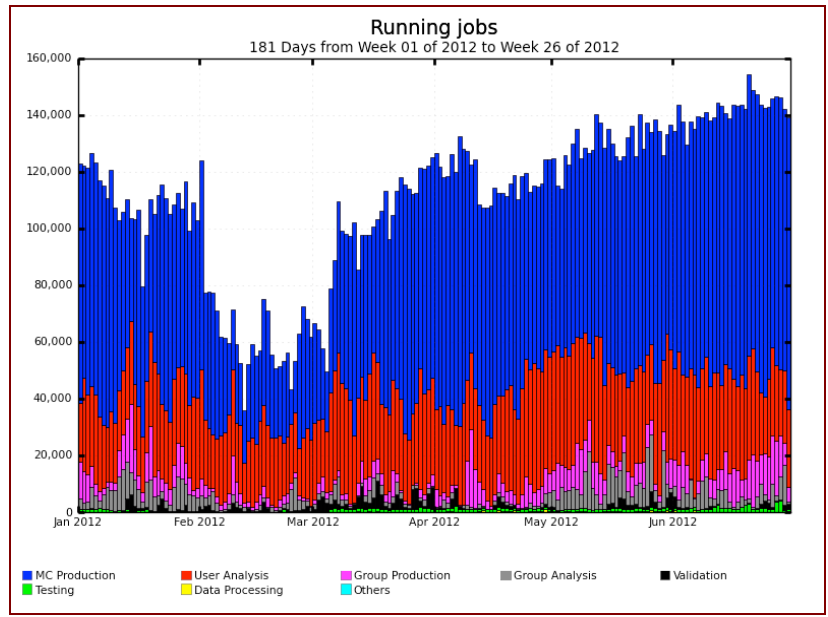

**Figure 2:** During first half of 2012, the number of concurrent Grid jobs in the ATLAS experiment routinely exceeded the level of 100,000. These jobs were running at the CERN Tier-0 site, ten large Tier-1 sites, and more than eighty smaller Tier-2 sites [1].

The observed signal significance is related to the LHC luminosity resulting in high data acquisition rates and petabytes of recorded data volumes, which requires significant computing power to process. Depending on conditions, it takes 3–6 $10^6$ core-hours to processes one petabyte of LHC data. Even higher computing power is required to produce simulated events required for the signal and background selection studies. Speakers at the CERN seminar acknowledged the role of Grid computing technologies in the discovery. It would have been impossible to release physics results so quickly without the outstanding performance of the Grid, including the CERN Tier-0 site. The ATLAS Grid resources were fully used. The number of running jobs often exceeded 100,000 including simulations, user analysis and group production (Figure 2). Figure 3 shows simulation capabilities of Grid computing in the CMS experiment.

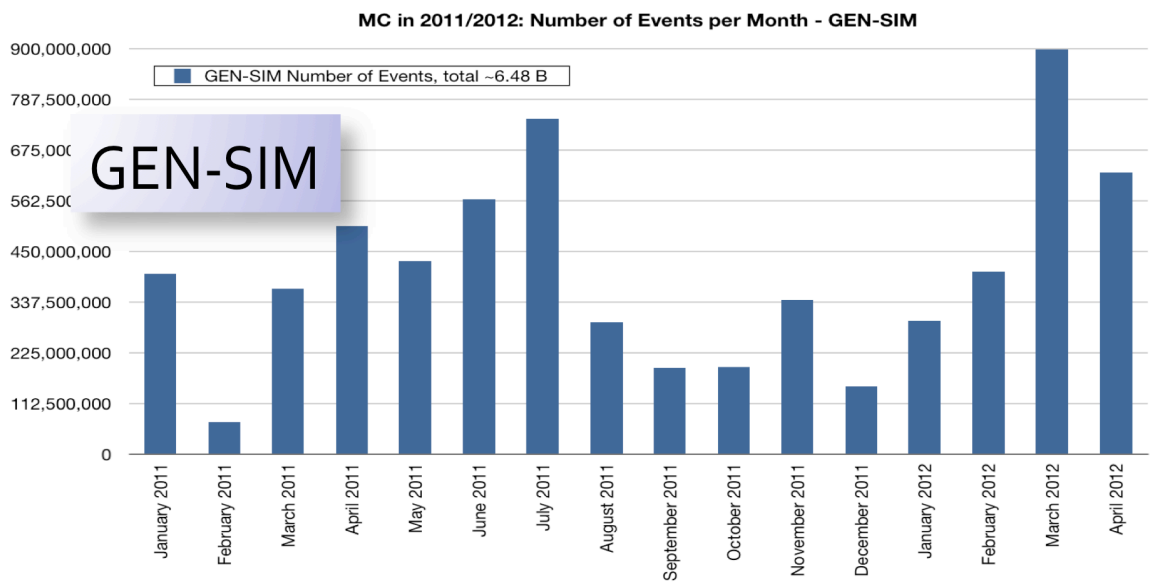

**Figure 3:** Thanks to the Grid computing capabilities, the CMS experiment achieved the sustained simulations rate at the level of 400,000,000 events per month [2].

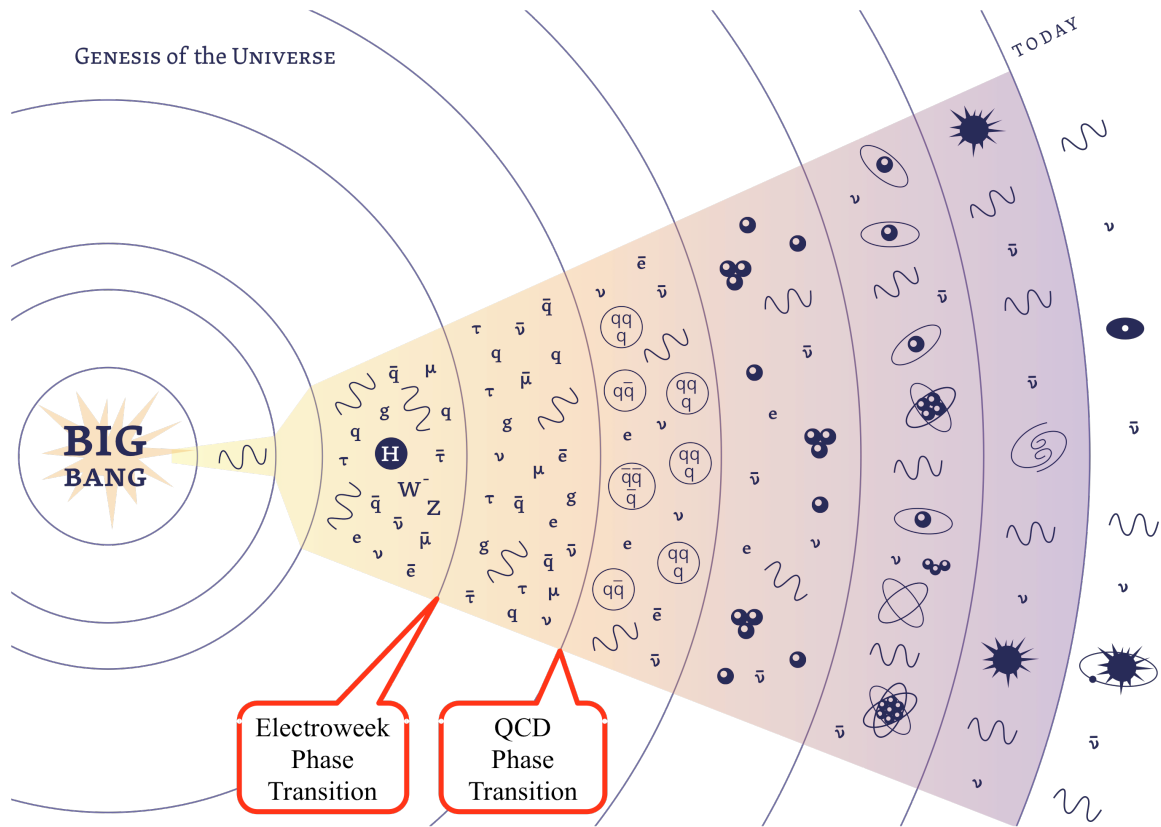

**Figure 4:** Evolution of the Universe [3].

## 4. Genesis

Figure 4 shows the time evolution of the Universe. As the temperature decreased with time, transitions took place. Among the Big Data experiments at LHC, the ALICE experiment studies the QCD phase transition; the ATLAS and CMS experiments probe the electroweak phase transition by observing the Higgs properties. It is possible that the matter-antimatter symmetry was broken earlier via the CP–violation mechanism [4]. However, the CP–violation in the Standard Model is too small for baryogenesis indicating new physics beyond the Standard Model. Future experiments in search for new physics (SuperB and Belle II) require Big Data technologies. In a high rate scenario, the Belle II experiment acquires data at the rate of 1,800 MB/s – at the level expected for all LHC experiments combined [5]. The storage needs grows from 50 PB to 600 PB in six years of the SuperB experiment [6]. Both experiments adopted Grid computing for Big Data processing [7, 8].

## 5. Six Sigma Quality for Big Data

In industry, Six Sigma analysis improves the quality of production by identifying and removing the causes of defects. A Six Sigma process is one in which products are free of defects at $0.3 \cdot 10^{-5}$ level because an industrial Six Sigma process corresponds to the mathematical $4.5\sigma$ after taking into account the $1.5\sigma$ shift from variations in production.

In contrast, LHC Big Data processing achieves $6\sigma$ quality in a true mathematical sense – the $10^{-8}$ level of defects. Figure 5 shows why physics requires $6\sigma$ quality during Big Data processing. In comparison to known physics processes, the production rate of new phenomena is very small. To select interesting data, LHC experiments employ hardwired multi-level data selection mechanisms (online trigger) followed by flexible offline selections (offline data processing). The production rates for Higgs physics at the LHC energies result in low selection rates – few events are selected out of billions recorded; experiments cannot loose even one event. Figure 6 shows number of events (after all selections) in the "golden" Higgs discovery channel illustrating that the event selection at the LHC is indeed at the $10^{-9}$ level [10].

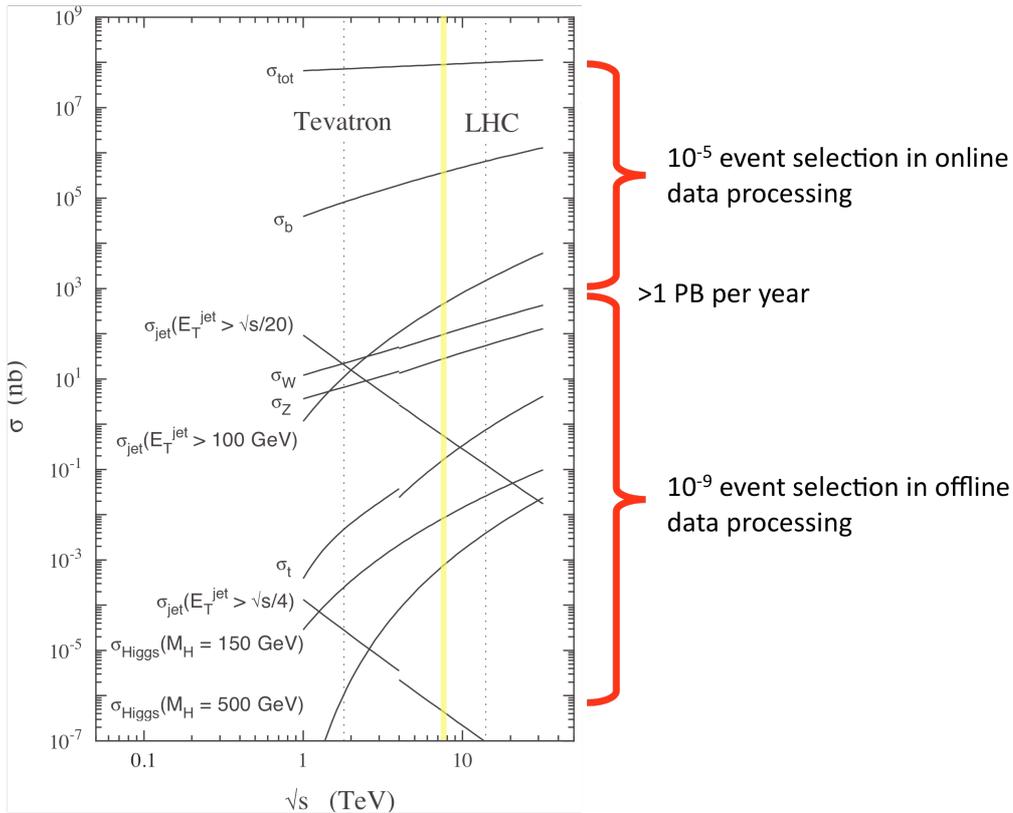

**Figure 5:** The offline event selection rate in LHC physics exploration [9] requires 6σ quality during Big Data processing at LHC energies shown by the yellow band.

Failure recovery by re-tries achieves production quality in Big Data processing on the Grid at the 6σ level [11]. No events were lost during the main ATLAS reprocessing campaign of the 2010 data that reconstructed on the Grid more than 1 PB of data with $0.9 \cdot 10^9$ events. In the last 2011 data reprocessing, only two collision events out of $0.9 \cdot 10^9$ events total could not be reconstructed. (These events were reprocessed later in a dedicated data recovery step.)

Later, silent data corruption was detected in six events from the reprocessed 2010 data and in one case of five adjacent events from the 2011 reprocessed data [11]. Corresponding to event losses below the $10^{-8}$ level, this demonstrates sustained 6σ quality performance in Big Data processing.

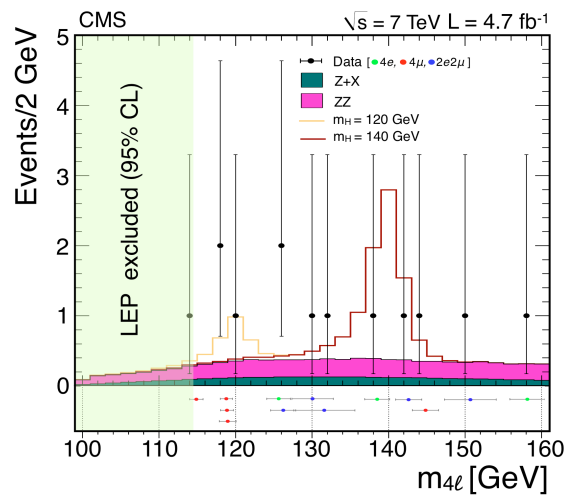

**Figure 6:** The event selection rate in the Higgs to four leptons channel [10]. Few events were selected out of a billion.

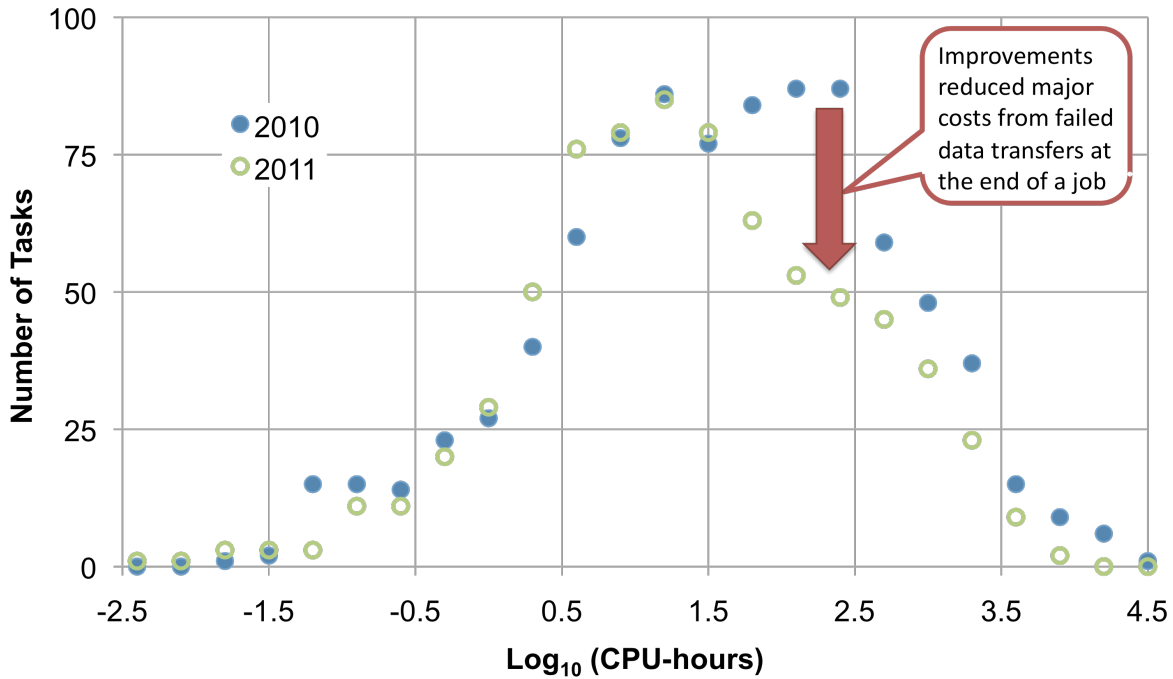

**Figure 7:** Distribution of tasks vs. CPU-hours used to recover job failures follows a multi-mode Weibull distribution.

## 6. Big Data Processing Techniques

In Big Data processing scientists deal with datasets, not individual files. Thus, a "task" – not a "job" – is a major unit Big Data processing on the Grid. Splitting of a large data processing task into jobs is similar to the splitting of a large file into smaller TCP/IP packets during the FTP data transfer. Splitting data into smaller pieces achieves reliability by re-sending of the dropped TCP/IP packets. Likewise, in Big Data processing transient job failures are recovered by re-tries. In file transfer, the TCP/IP packet is a unit of checkpointing. In high energy physics Big Data processing, the checkpointing unit is a job (e.g., PanDA [12]) or a file (e.g., DIRAC [13]).

Many high-performance computing problems are tightly-coupled and require inter-process communications to be parallelized. In contrast, high energy physics computing often called embarrassingly parallel, since the units of data processing – physics events – are independent. However, the event-level checkpointing rarely used today, as its granularity is too small for Big Data processing in high energy physics. The next generation system needs the event-level checkpointing for Big Data.

## 7. Reliability Engineering for ATLAS Big Data Processing on the Grid

LHC computing experience has shown that Grid failures can occur for a variety of reasons. Grid heterogeneity makes failures hard to diagnose and repair quickly. Big Data processing on the Grid must tolerate a continuous stream of failures, errors and faults. The failure detection and performance prediction are considered open areas of research by many [14].

While fault-tolerance mechanisms improve the reliability of Big Data processing in the Grid, their benefits come at costs. Reliability Engineering provides a framework for fundamental understanding of the Big Data processing on the Grid, which is not a desirable enhancement but a necessary requirement.

*7.1. Failure Recovery Cost*

Job resubmission avoids data loss at the expense of CPU time used by the failed jobs. In 2010 reprocessing, the CPU time used to recover transient failures was 6% of the CPU time used for the reconstruction. In 2011 reprocessing, the CPU time used to recover transient failures was reduced to 4% of the CPU time used for the reconstruction. Figure 7 shows that most of the improvement came from reduction in failures in data transfers at the end of a job.

*7.3. Time Overhead*

Fault-tolerance achieved through automatic re-tries of the failed jobs induces a time overhead in the task completion, which is difficult to predict. Transient job failures and re-tries delay the reprocessing duration. Workflow optimization in ATLAS Big Data processing on the Grid and other improvements cut the delays and halved the duration of the petabyte-scale reprocessing on the Grid from almost two months in 2010 to less than four weeks in 2011 [11]. Optimization of fault-tolerance techniques to speed up the completion of thousands of interconnected tasks on the Grid is an active area of research in ATLAS.

**8. Summary**

The emerging Big Data revolution drives new discoveries in scientific fields including nanotechnology, astrophysics, high-energy physics, biology and medicine. In Big Data processing on the Grid, physicists deal with datasets, not individual files. A task (comprised of many jobs) became a unit of Big Data processing. Reliability Engineering provides a framework for fundamental understanding of Big Data processing on the Grid, which is not a desirable enhancement but a necessary requirement. Fault-tolerance achieved through automatic re-tries of the failed jobs induces a time overhead in the task completion, which is difficult to predict. Reduction of the duration of Big Data processing tasks on the Grid is an active area of research in ATLAS.

**Acknowledgements**

I wish to thank the Conference organizers for their invitation and hospitality. I also thank all my collaborators and colleagues who provided materials for this review. This work supported in part by the U.S. Department of Energy, Division of High Energy Physics, under Contract DE-AC02-06CH11357.